\documentclass[12pt,english]{article}
\usepackage[T1]{fontenc}
\usepackage[latin9]{inputenc}
\usepackage{geometry}
\usepackage{babel}
\usepackage{float}
\usepackage{amsmath}
\usepackage{graphicx}
\usepackage{setspace}
\usepackage[authoryear]{natbib}
\usepackage[unicode=true,pdfusetitle,
 bookmarks=true,bookmarksnumbered=false,bookmarksopen=false,
 breaklinks=false,pdfborder={0 0 0},pdfborderstyle={},backref=false,colorlinks=false]
 {hyperref}

\makeatletter
\usepackage{setspace}
\usepackage{lineno}
\usepackage{natbib}
\usepackage{pdflscape}
\usepackage{hyperref}
\usepackage{multirow}
\usepackage{xcolor}

\usepackage{titlesec}
\titlespacing*{\subsubsection}{0pt}{0.5ex}{0pt}
\titlespacing*{\subsection}{0pt}{1.0ex}{0pt}
\titlespacing*{\subsubsection}{0pt}{1.0ex}{0pt}
\titlespacing*{\subsection}{0pt}{1.0ex}{0pt}

\makeatother

\begin{document}
\begin{flushleft}
\setcounter{footnote}{0}
\begin{flushleft}\textbf{When and where: estimating the date and location of introduction
for exotic pests and pathogens}
\par\end{flushleft}

\medskip{}
\medskip{}
\medskip{}
\begin{spacing}{1}

\begin{singlespace}
\textbf{Trevor J. Hefley}\footnote{Corresponding author: thefley@ksu.edu\\
}

{\small{}Department of Statistics}{\small\par}

{\small{}Kansas State University}{\small\par}
\end{singlespace}

\begin{doublespace}
\medskip{}

\end{doublespace}

\begin{singlespace}
\textbf{Robin E. Russell}

U.S. Geological Survey

National Wildlife Health Center
\end{singlespace}

\begin{doublespace}
\medskip{}

\end{doublespace}

\begin{singlespace}
\textbf{Anne E. Ballmann}

U.S. Geological Survey

National Wildlife Health Center
\end{singlespace}

\begin{doublespace}
\medskip{}

\end{doublespace}

\begin{singlespace}
\textbf{Haoyu Zhang}

{\small{}Department of Statistics}{\small\par}

{\small{}Kansas State University}{\small\par}
\end{singlespace}

\bigskip{}
\medskip{}
\medskip{}

\setlength{\parindent}{0.7cm}

A fundamental question during the outbreak of a novel disease or invasion
of an exotic pest is: At what location and date was it first introduced?
With this information, future introductions can be anticipated and
perhaps avoided. Point process models are commonly used for mapping
species distribution and disease occurrence. If the time and location
of introductions were known, then point process models could be used
to map and understand the factors that influence introductions; however,
rarely is the process of introduction directly observed. We propose
embedding a point process within hierarchical Bayesian models commonly
used to understand the spatio-temporal dynamics of invasion. Including
a point process within a hierarchical Bayesian model enables inference
regarding the location and date of introduction from indirect observation
of the process such as species or disease occurrence records. We illustrate
our approach using disease surveillance data collected to monitor
white-nose syndrome, which is a fungal disease that threatens many
North American species of bats. We use our model and surveillance
data to estimate the location and date that the pathogen was introduced
into the United States. Finally, we compare forecasts from our model
to forecasts obtained from state-of-the-art regression-based statistical
and machine learning methods. Our results show that the pathogen causing
white-nose syndrome was most likely introduced into the United States
4 years prior to the first detection, but there is a moderate level
of uncertainty in this estimate. The location of introduction could
be up to 510 km east of the location of first discovery, but our results
indicate that there is a relatively high probability the location
of first detection could be the location of introduction.

\section{Introduction}

When a species or pathogen invades a novel environment, determining
when and where it was first introduced is of interest to scientists,
policymakers, and the public. Knowing the location and date of introduction
would further our understanding of the ecology of the process, increase
our capacity to build predictive models, and could inform policies
that aim to prevent new introductions. Although the location and time
of introduction is known in some cases, such as the introduction of
myxomytosis to control rabbits in Australia (\citealt{ratcliffe1952myxomatosis}),
there are many examples where the location and time of introduction
are unknown and must be inferred from indirect data.

The general problem of modeling the spatio-temporal dynamics of invasion
is well-represented in the literature (e.g., \citealt{mollison1986modelling,gibson1996fitting,wikle2003invasion});
however, when the introduction is not directly observed there is a
small number of limited methods that might be used to determine the
location and time of introduction from indirect data. For example,
a method known as geographic profiling has been introduced to estimate
the position of sources of invasive species and diseases using occurrence
records (\citealt{le2011geographic,stevenson2012geographic,verity2014spatial}),
but this approach ignores the temporal dynamics and therefore cannot
estimate the initial source (however see \citealt{mohler2012geographic}).
Similarly, Bayesian models have been developed to estimate the spatio-temporal
dynamics of colonization (e.g., \citealt{cook2007bayesian,catterall2012accounting,broms2016dynamic}),
but these methods are limited. For example, the \citet{cook2007bayesian}
approach requires that the colonization time be known. \citet{catterall2012accounting}
demonstrate how to infer the time of colonization; however, this method
was developed to understand the dynamics after the initial introduction
and not to obtain estimates of the time and location of the first
introduction. Lastly, \citet{caley2015inferring} developed a Bayesian
spatio-temporal model to infer the time of introduction and geographic
distribution of an invasive species from occurrence records; however,
this approach does not provide a framework for inferring the location
of introduction.

If the time and location of introductions were known, then tools commonly
used to map the distribution of species and diseases could be used
to build predictive models and infer the factors that influence the
occurrence of novel introductions. Novel introductions, however, are
rarely directly observed. Instead, the outcome of a successful introduction
is observed by the collection of indirect data such as species or
disease occurrence records. In principal, indirect spatio-temporal
data and models can be used infer historical events such as infection
histories (\citealt{salje2018reconstruction}), changes in climate
(\citealt{tingley2012piecing,tipton2016reconstruction}), and, as
we demonstrate, introductions of pests and pathogens.

Our objective is to show how point process models, a commonly used
statistical approach for modeling the distribution of species and
diseases, can be placed within a hierarchical modeling framework to
obtain statistically principled estimates of the date and location
of introduction. Although we provide a general framework, our work
is motivated by the need to understand when and where $\textit{Pseudogymnoascus destructans}$,
the pathogen that causes white-nose syndrome (WNS), was first introduced
into the United States. White-nose syndrome was first detected in
2006 using photographic evidence of a bat from Howes Cave, near Albany,
New York (Fig. 1; \citealt{blehert2009bat,frick2010emerging}). The
pathogen, $\textit{P. destructans}$, has since spread throughout
the eastern and midwestern United States resulting in high mortality
rates among several species of cave-hibernating bats. Statistical
methods capable of estimating the date and location where $\textit{P. destructans}$
was first introduced into the United States could further our understanding
of the etiology of the disease and provide critical information on
the process of introduction. For example, WNS was recently detected
on a western subspecies of little brown bat (\textit{Myotis lucifugus})
occurring in the state of Washington that is likely the result of
a novel introduction originating from the eastern United States (\citealt{lorch2016first}).
Our modeling approach could eventually be used to test the hypothesis
that the Washington WNS case is the result of a novel introduction
and, if confirmed, identify areas at high risk for future introductions.
\begin{figure}[H]
\begin{centering}
\includegraphics[scale=0.7]{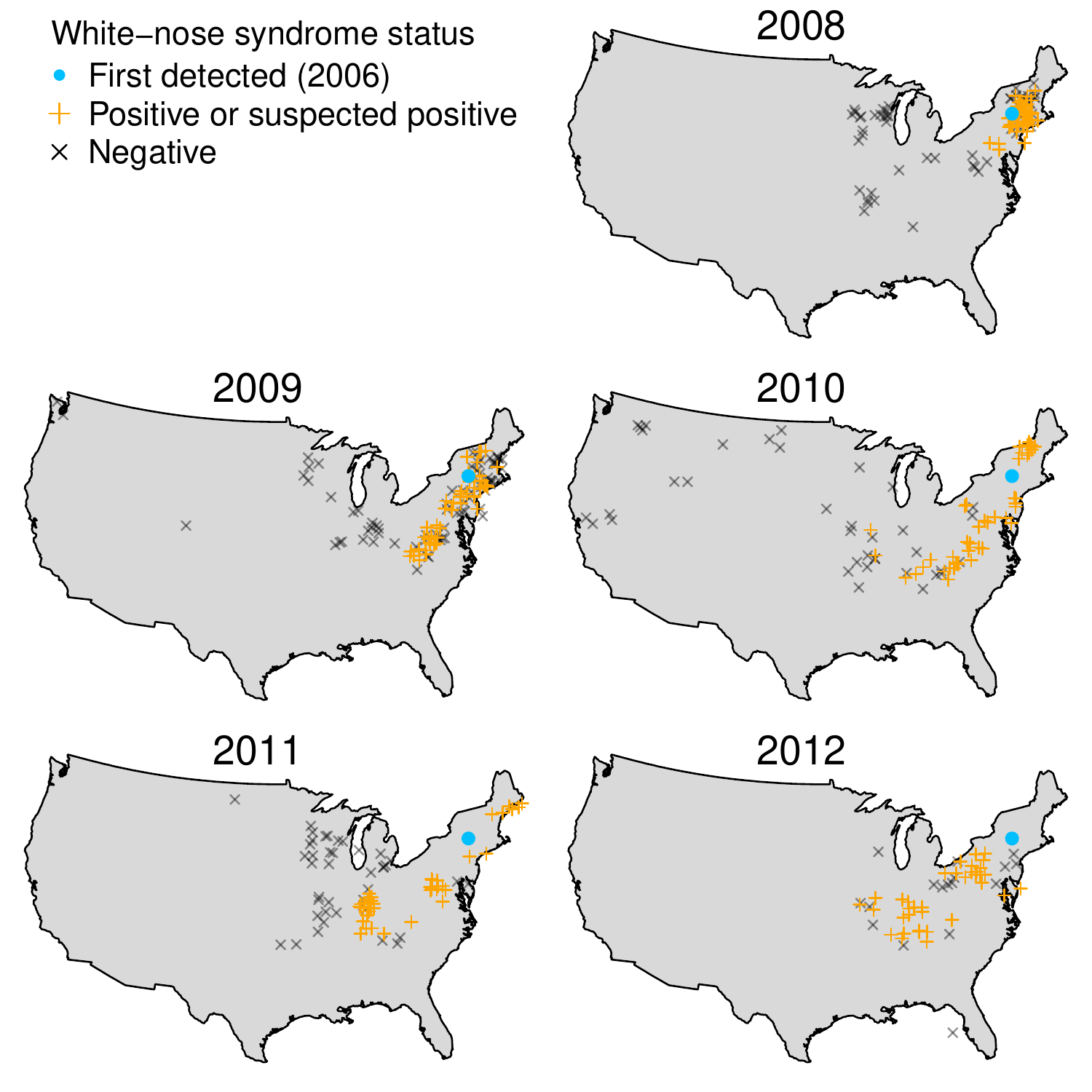}
\par\end{centering}
\noindent \textbf{Fig. 1.} Location and infection status of individual
bat samples from four species tested for white-nose syndrome. The
first detection in 2006 was based on photographic evidence only. The
locations of individual samples were jittered so that cases from the
same location did not overlap. The four species of bats tested include:
little brown bat (\textit{Myotis lucifugus}), big brown bat (\textit{Eptesicus
fuscus}), northern long-eared bat (\textit{Myotis septentrionalis}),
and tri-colored bat (\textit{Perimyotis subflavus}).
\end{figure}

\section{Dynamic Spatio-temporal Statistical Models}

Dynamic spatio-temporal statistical models are used in fields such
as ecology (\citealt{wikle2003invasion,hooten2008doves,williams2017integrated,hefley2017mechanism}),
epidemiology (\citealt{smith2002predicting}), and meteorology (\citealt{wikle1998hierarchical,wikle2001spatiotemporal}).
The defining property of a dynamic statistical model is an explicit
description of how future observations (spatial data) evolve from
past observations (\citealt{wikle2010general,mohler2012geographic,cressie2011statistics,wikle2018stwr}).
For example, statistical implementations of dynamic agent-based models
are constructed using a set of deterministic or stochastic rules governing
the actions of individual agents at each time step (\citealt{hooten2010agent,hooten2010statistical,keith2013agent,heard2015agent,mcdermott2017hierarchical}).

Consider the classic example of a dynamic agent-based model where
an individual performs a random walk. At each time step, the individual
moves a random distance in a random direction. An inherent characteristic
of dynamic models is that the initial conditions must be specified.
For example, when performing a random walk, the location and time
the individual begins walking must be specified. Although a Eulerian
(individual-based) frame of reference can be used to understand dynamic
processes, a Lagrangian (population-level) perspective is also common
(\citealt{hooten2017animal}). In ecology, for example, dynamic models
of populations are expressed as partial differential equations (PDEs),
which are used to describe the global behavior of a large number of
individuals (\citealt{holmes1994partial,hooten2017animal}).

In data-driven modeling applications, hierarchical (conditional) model
specifications can be used to estimate unknown parameters by connecting
data to the underlying dynamic process (\citealt{cressie2011statistics,wikle2018stwr}).
Many applications of dynamic statistical models have focused on estimating
parameters that govern the process (e.g., rate of spread), but rarely
has formal statistical inference about the initial conditions been
of interest (but see \citealt{hefley2017spatial}). Instead, the initial
conditions are modeled phenomenologically and treated as nuisance
parameters (\citealt{wikle2003hierarchical,hooten2008doves,williams2017integrated,hefley2017mechanism}).

We demonstrate how incorporating scientific knowledge about the initial
conditions of a dynamic process provides novel insight into the outbreak
of an infectious disease or invasion of a pest. By justifiably assuming
that the pathogen or pest was introduced at points occurring at unknown
locations and times, we specify a point process to model the initial
conditions. Point process models are commonly used for species distribution
and disease mapping (\citealt{renner2015point}), but can also be
incorporated into hierarchical Bayesian models to learn about latent
(unobserved) processes (e.g., \citealt{BB2L}). In the following sections,
we provide a specific example using a PDE of ecological diffusion;
however, our approach is applicable to commonly used dynamic models
that require the specification of initial conditions such as agent-based
models, contact process models, rule-based approaches (e.g., cellular
automaton), integral equations, and integro-difference equations.

\subsection{Ecological Diffusion}

Diffusion is a dynamic process that can be used to describe the movement
of particles, animals, pathogens, or other objects from a Lagrangian
perspective (\citealt{hooten2017animal}). A specific type of diffusion,
called ecological diffusion (\citealt{hooten2017animal}), has been
used to explain the spread of disease (\citealt{garlick2011homogenization,garlick2014homogenization,hefley2017spatial,hefley2017mechanism}),
the outbreak of pests (\citealt{powell2014phenology,hooten2013computationally,powell2017differential}),
the invasions of species (\citealt{neupane2015invasion}), and the
reintroduction of extirpated species (\citealt{williams2017integrated}).
Ecological diffusion is expressed by the PDE 
\begin{equation}
\frac{\partial}{\partial t}u(\mathbf{s},t)=\left(\frac{\partial^{2}}{\partial s_{1}^{2}}+\frac{\partial^{2}}{\partial s_{2}^{2}}\right)[\mu(\mathbf{s},t)u(\mathbf{s},t)]
\end{equation}

\noindent where $u(\mathbf{s},t)$ is the intensity of the dispersing
pathogen (or population), $s_{1}$ and $s_{2}$ are the spatial coordinates
contained in the vector $\mathbf{s}$, and $t$ is the time. The intensity,
$u(\mathbf{s},t)$, can be linked to the expected pathogen abundance
by integrating $u(\mathbf{s},t)$ over a spatial domain of interest
(i.e., $\int_{\mathcal{A}}u(\mathbf{s},t)d\mathbf{s}$). The diffusion
coefficient, $\mu(\mathbf{s},t)$ could depend on location-specific
covariates that control the rate of spread and vary over time. Although
equation (1) only describes ecological diffusion, PDEs can be modified
to capture other important components such a population growth or
long-range dispersal (\citealt{holmes1994partial}). For example,
in our WNS data example we modify equation (1) to include an exponential
growth component to enable the pathogen $\textit{P. destructans}$
to increase in abundance.

Ecological diffusion describes how the spatial distribution of a pathogen
or population evolves over time. To initiate this dynamic process,
the conditions at the time of introduction, $t_{0}$, must be known
or estimated. In situations of disease outbreak neither $t_{0}$ or
$u(\mathbf{s},t_{0})$ are known. A realistic assumption is to presume
that the pathogen was introduced at points, which can be formulated
mathematically as
\begin{equation}
u(\mathbf{s},t_{0})=\begin{cases}
\theta_{j} & \text{if}\;\mathbf{s}=\boldsymbol{\omega}_{j}\\
0 & \text{if}\;\mathbf{s}\neq\boldsymbol{\omega}_{j}
\end{cases}\:,
\end{equation}

\noindent where $\theta_{j}$ is the unknown initial number of pathogen
particles (or animals) and $\boldsymbol{\omega}_{j}$ is the unknown
coordinate of the $j^{\text{th}}$ location of introduction ($j=1,2,\ldots,J$).
In data-driven applications, the time, location, initial number of
particles, and potentially the number of points are unknown, thus
the parameters $t_{0}$, $\boldsymbol{\omega}_{j}$, $\theta_{j}$,
and $J$ must be estimated.

Estimation of the initial conditions can be understood heuristically
as follows. By observing data generated from a diffusion process at
multiple time points (Fig. 2), it is feasible to estimate the rate
at which the process is diffusing (spreading). Once the diffusion
rate is estimated, equation (1) can predict the future (forecast)
or past (backcast) spread of the pathogen. Unlike forecasting, backcasting
ultimately results in a highly constrained state with a functional
form that we have knowledge about. Specifically, if the diffusion
processes started at points, then backcasting has a known endpoint
that constrains the functional form of $u(\mathbf{s},t_{0})$ to equation
(2). This constraint, formulated as a prior or process-level distribution
when using a hierarchical Bayesian statistical model, enables the
posterior distribution of the time, location, and number of points
to be calculated from data. Next we introduce the data set that motivates
our work. Understanding the specifics of the data is required to specify
an appropriate hierarchical Bayesian model that links the ecological
diffusion PDE to observed data and enables us to estimate the date
and location of introduction.
\begin{figure}[H]
\begin{centering}
\includegraphics[scale=0.6]{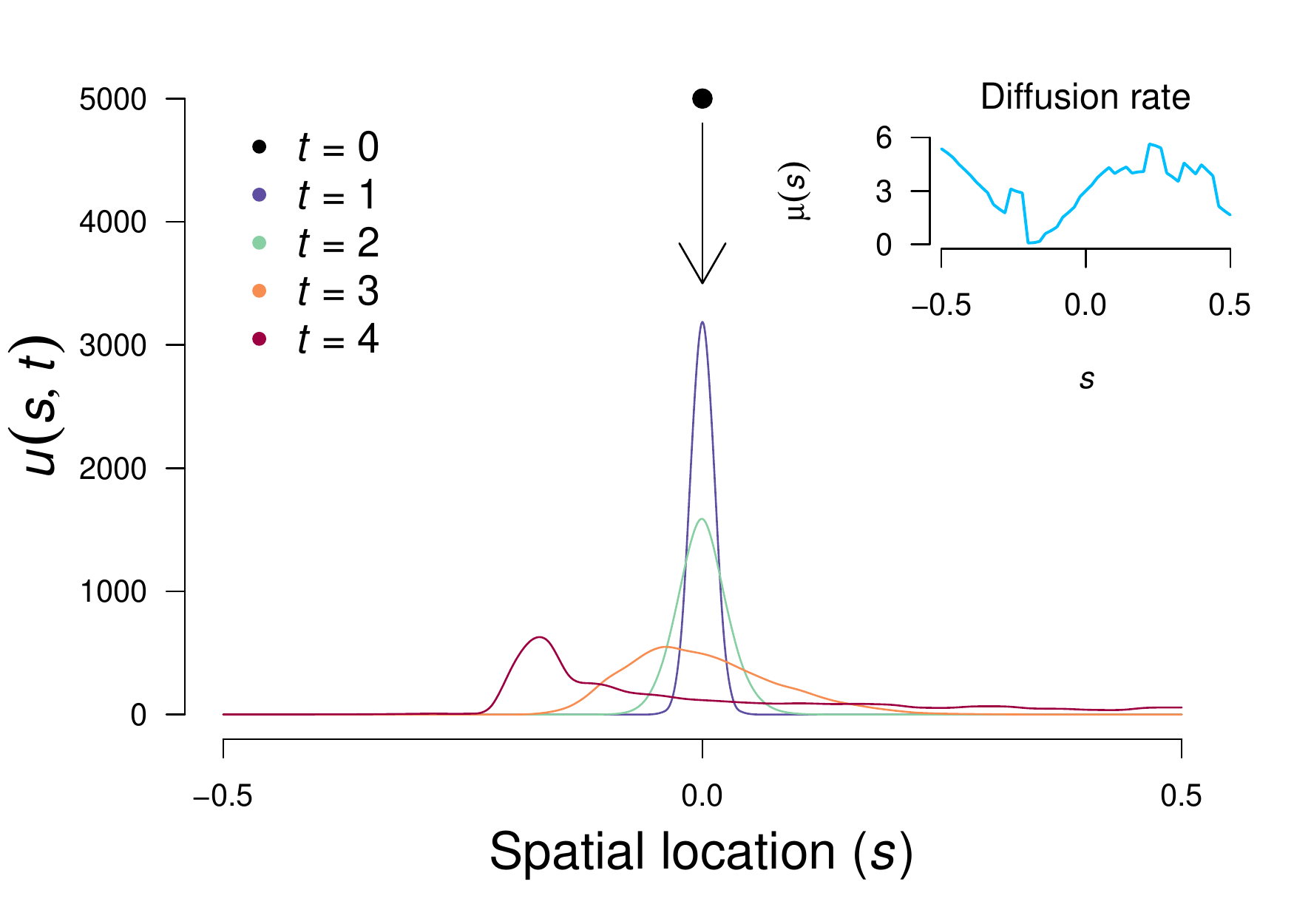}
\par\end{centering}
\noindent \textbf{Fig. 2.} Illustration showing an initial introduction
of $\theta=100$ particles and how the one-dimensional spatial distribution
of pathogen or population intensity ($u(s,t)$) evolves over time
according to the ecological diffusion partial differential equation.
Greater values of the diffusion rate $\mu(s)$, shown in the inset
plot, result in faster spread whereas spread is inhibited when $\mu(s)=0$.
The diffusion rate depends on the environmental variables at the location
($s$), which results in the spatial variability (roughness) visible
in $u(s,t)$. The pathogen intensity, $u(s,t)$, can be linked to
the expected pathogen abundance by integrating $u(s,t)$ over an interval
of interest (i.e., $\int_{a}^{b}u(s,t)ds$). In this example, $\int_{-0.5}^{0.5}u(s,t)ds=100$
for all time points.
\end{figure}

\section{White-nose Syndrome Data Example}

Surveillance for WNS in the United States began in 2007 using a combination
of passive and active surveillance methods. During 2007\textendash 2012,
samples were obtained from individual bats associated with morbidity
or mortality investigations occurring year-round at underground hibernacula
or on the above-ground landscape, and consisted of bat carcasses,
biopsies of wing skin, or tape lifts of fungal growth on muzzles.
A small number of samples were also obtained from target species (including
$\textit{Myotis}$ spp., $\textit{Perimyotis subflavus}$, $\textit{Eptesicus fuscus}$)
admitted to rehabilitation facilities or state diagnostic laboratories
for rabies testing from approximately December to May. A positive
or negative diagnosis of WNS in individual bats was determined by
observing characteristic histopathologic lesions in skin tissues using
light microscopy (\citealt{meteyer2009histopathologic}). A diagnosis
of suspect WNS was assigned to individuals with clinical signs suggestive
of the disease that either had ambiguous skin histopathology or were
not evaluated in this manner but for which the causative agent, $\textit{P. destructans}$,
was detected by fungal culture or polymerase chain reaction (\citealt{lorch2010rapid}).

We illustrate our modeling approach using a subset of the WNS surveillance
data collected during 2008\textendash 2011 that includes individual
samples of little brown bats (\textit{Myotis lucifugus}), big brown
bats (\textit{Eptesicus fuscus}), northern long-eared bats (\textit{Myotis
septentrionalis}), and tri-colored bats (\textit{Perimyotis subflavus}).
This resulted in $\geq50$ individual samples for each species and
a total of 397 samples with 184 positive or suspected positive cases
of WNS (Fig. 1). We limited our illustration to the four most common
species sampled because the remaining species were uncommon and included
less than 20 individuals and no more than five positive cases for
each species. In our analysis, we chose to exclude the first case
detected in 2006 because it was based on photographic evidence as
well as samples from 2007 because diagnostic case criteria for WNS
were not established until 2008. Finally, we used data collected from
2012 for the same four species to test the ability of our model to
forecast future spread of $\textit{P. destructans}$ and prevalence
of WNS. This resulted in $\geq3$ individuals for each species and
a total of 54 individuals with 37 positive or suspected positive cases.

\subsection{Hierarchical Bayesian Statistical Model}

We take a hierarchical Bayesian approach and specify a statistical
model that facilitates simultaneous parameter estimation of the location
and date of introduction as well as the diffusion and growth rate
(\citealt{hooten2013computationally}). Our model is tailored to the
WNS data (Fig. 1), but the hierarchical Bayesian approach is flexible
and can be adapted to a wide range of studies (e.g., \citealt{Hobbs2015,williams2017integrated}).
For the WNS data, we specify the hierarchical model:\begin{eqnarray}
y_{i} & \sim & \text{Bernoulli}\left(p_{i}\right)\\
p_{i} & = & g^{-1}(u(\mathbf{s}_{i},t_{i})e^{\mathbf{x}^{\prime}_{i}\boldsymbol{\beta}})\\ 
\frac{\partial}{\partial t}u(\mathbf{s},t) & = & \left(\frac{\partial^{2}}{\partial s_{1}^{2}}+\frac{\partial^{2}}{\partial s_{2}^{2}}\right)[\mu(\mathbf{s})u(\mathbf{s},t)]+\lambda(\mathbf{s})u(\mathbf{s},t)\\ 
\text{log}\left(\mu(\mathbf{s})\right) & = & \alpha_{0}+\mathbf{z}(\mathbf{s})^{\prime}\mathbf{\boldsymbol{\alpha}}\\ 
\lambda(\mathbf{s}) & = & \gamma_{0}+\mathbf{w}(\mathbf{s})^{\prime}\mathbf{\boldsymbol{\gamma}},
\end{eqnarray}where $y_{i}$ is equal to $1$ if the $i^{\text{th}}$ tested bat
is WNS-positive or suspected positive and $0$ if negative. The probability
that $y_{i}=1$, $p_{i}$, depends on the pathogen intensity $u(\mathbf{s}_{i},t_{i})$
defined by the PDE in equation (5) and species susceptibility $e^{\mathbf{x}_{i}^{\prime}\boldsymbol{\beta}}$,
where $\mathbf{x}_{i}$ is a vector that contains a binary indicator
variable identifying the species of the $i^{\text{th}}$ tested bat
and $\boldsymbol{\beta}\equiv(\beta_{1},\ldots,\beta_{p_{\beta}})^{'}$
is a vector of species-specific susceptibility coefficients. Differences
among species susceptibility to WNS has been reported although the
mechanism(s) for these differences remain unclear (\citealt{langwig2012sociality,johnson2015antibodies,langwig2015invasion}).
Regardless of the mechanism(s), our model accounts for heterogeneity
among species susceptibility by inclusions of $e^{\mathbf{x}_{i}^{\prime}\boldsymbol{\beta}}$.

For the ecological diffusion PDE, the product of the pathogen intensity
($u(\mathbf{s}_{i},t_{i})$) and species susceptibility ($e^{\mathbf{x}_{i}^{\prime}\boldsymbol{\beta}}$)
in equation (4) is greater than zero and therefore the inverse link
function $g^{-1}(\cdot)$ maps the positive real line $[0,\infty)$
to a probability between 0 and 1. For an inverse link function, we
used the cumulative distribution function of a standard log-normal
distribution (i.e., $g^{-1}(x)=\frac{1}{x\sqrt{2\pi}}\int_{0}^{x}e^{-\frac{1}{2}\text{log}(x)^{2}}dx$).
Although any appropriate link function could be used, our choice enables
efficient implementation because the full-conditional distribution
for $\boldsymbol{\beta}$ are available in closed-form (see Appendix
S1 for details; \citealt{BB2L}). The diffusion rate ($\mu(\mathbf{s}$))
in equation (5) depends on the regression-type equation in (6) that
has an intercept term ($\alpha_{0}$), coefficients ($\boldsymbol{\alpha}$$\equiv(\alpha_{1},\ldots,\alpha_{p_{\alpha}})^{'}$),
and location-specific covariates $\mathbf{z}(\mathbf{s})$. For spreading
diseases, the diffusion rate is always positive, thus $\mu(\mathbf{s})>0$
for all $\mathbf{s}$, which motivates the log link function in (6).
In addition to ecological diffusion, we add an exponential growth
component to the PDE to account for increasing (or decreasing) pathogen
intensity. This addition results in a growth rate ($\lambda(\mathbf{s})$)
that depends on a similar regression-type equation (7) that has an
intercept term ($\gamma_{0}$), coefficients ($\boldsymbol{\gamma}\equiv(\gamma_{1},\ldots,\gamma_{p_{\gamma}})^{\prime}$),
and location-specific covariates $\mathbf{w}(\mathbf{s})$, which
can be the same as or different from $\mathbf{z}(\mathbf{s})$. To
allow the growth rate to be positive or negative, we used the identity
link function so that the support of $\lambda(\mathbf{s})$ is unconstrained
and can take on values from $-\infty$ to $\infty$.

We assume Dirichlet boundary conditions and set $u(\mathbf{s},t)=0$
at the boundary of the United States. Similar to initial conditions,
boundary conditions could be an important component and future studies
may need to estimate the boundary conditions. For computational efficiency,
we prefer Dirichlet boundary conditions because existing techniques
for estimation would require estimating a large number of parameters
(\citealt{wikle2003hierarchical}).

To fully specify a hierarchical Bayesian model, prior distributions
must be carefully chosen for all parameters. We describe and justify
our choice of prior distributions in Appendix S2, with the exception
of the prior distribution for the location of introduction, which
we describe below. We specify the initial conditions as described
in equation (2). The natural choice for the prior of the latent (unobserved)
locations of introduction is a point process, which is a probability
distribution that generates locations in geographic space as random
variables (\citealt{cressie2011statistics}). For the WNS data, we
assume a single point source introduction (i.e., $J=1$), which results
in a conditional point process (i.e., a point process with a known
number of points, in this case just one; \citealt{cressie1993statistics}
p. 651). For an introduction that occurs at a single location, the
intensity function of the point process must be specified as a hyperparameter;
we specify the intensity function of the point process to be constant
across the United States which is equivalent to assuming that $\textit{P. destructans}$
was equally likely to have been introduced at any location within
the United States.

If multiple introductions occur, it may be feasible to estimate the
intensity function of the point process (Appendix S1). For example,
the intensity function may be modeled using a regression-type equation
to estimate how location-specific risk factors influence the likelihood
of novel introductions. This approach could then be used to predict
the geographic areas and identify the environmental conditions where
new introductions are likely to occur. Finally, our approach can be
modified to handle an unknown number of point sources that are introduced
on different dates by using a spatio-temporal Poisson point process
(Appendix S1). This modification may eventually be used to explain
the recent positive cases of WNS in the state of Washington (\citealt{lorch2016first}).

\subsection{Model Implementation}

The covariates $\mathbf{z}(\mathbf{s})$ and $\mathbf{w}(\mathbf{s})$,
which influence the diffusion and growth rate in equations (6) and
(7), respectively, can include any location-specific environmental
factor (e.g., temperature). We used karst terrain and deciduous forest
cover calculated from digitized maps (\citealt{tobin2004digital})
and the 2011 National Land Cover Database (\citealt{NLCD2011}), respectively
(Fig. 3). We expect that the spatial availability of caves and deciduous
forest habitat will influence the diffusion of the pathogen due to
the ability of bats to disperse; however, we acknowledge many other
factors could be included depending on the goals of the study.
\begin{figure}[H]
\begin{centering}
\includegraphics{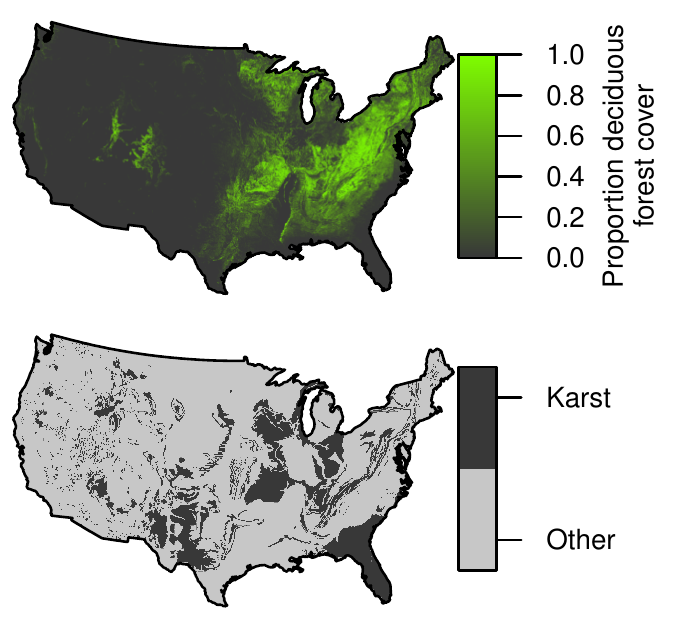}
\par\end{centering}
\noindent \textbf{Fig. 3.} The Environmental covariates proportion
of deciduous forest (top panel) and karst terrain (bottom panel).
\end{figure}

For model implementation, we first use an analytical technique called
homogenization that results in a computationally efficient approximation
to the solution of the ecological diffusion PDE (\citealt{garlick2011homogenization,hooten2013computationally}).
To solve the homogenized PDE, we use a standard finite-difference
scheme and chose 30 time steps per month, broad-scale grid cells of
100 km by 100 km, and the fine-scale grid cells of 10 km by 10 km
(\citealt{Farlow1993PDE,garlick2011homogenization,hooten2013computationally}).
Finer spatial discretization may be implemented, but the resolution
we chose is a reasonable trade-off between the location accuracy of
our data (mostly reported at the county-level) and what is computationally
feasible using accessible high-performance computing. For implementation
of our Bayesian model, we used a Markov chain Monte Carlo (MCMC) algorithm
with three chains. For each chain, we obtained 320,000 samples from
the posterior distribution and discarded the first 20,000 samples
to ensure that the MCMC algorithm was sampling from the target (stationary)
distribution. All computations were conducted using a program in R
(Appendix S1 and S2) with optimized basic linear algebra subprograms
and subroutines written in C++ and called from R (\citealt{TeamR}).

\subsection{Model and Software Validation}

As noted by \citet{cook2006validation}, implementing non-standard
models such as ours often requires developing the necessary software
\textquotedblleft from scratch,\textquotedblright{} which may increase
the chance of making errors when programming. Furthermore, when using
a complex statistical model it is important to check that the parameters
are identifiable (\citealt{lele2010model}). One way to check our
software for errors and our proposed model for identifiability of
parameters is to use a simulated data set that mimics the WNS surveillance
data, but where the parameter values are known. If our software and
model are able to recover the known values of the parameters from
the simulated data that mimics the WNS surveillance data, then there
is evidence that the software is error free and that the parameters
in our model are identifiable. Appendix S1 contains a simulated data
example to demonstrate that our model is capable of recovering known
values of the parameters (see Figs. S1 \& S2).

In addition to a single simulated data set that mimics the WNS surveillance
data, we conducted a simulation experiment that involved simulating
several thousand data sets under different settings and fitting our
proposed model to each data set. A simulation experiment can be used
to evaluate frequentist properties of the estimated parameters, such
as bias and coverage probability, which ensures statistical inference
from our model is valid (\citealt{little2006calibrated}). We conducted
a simulation experiment employing a simplified version of the model
we used for WNS surveillance data, but included a setting for multiple
introductions (see Table S3 and Figs. S6\textendash S7 for simulation
settings in Appendix S3). For each setting, we evaluated bias of the
estimated time and location as well as the coverage probability of
the 90\% credible intervals (or regions) for each parameter. We expect,
unless something is pathologically wrong with our model (e.g., unidentifiable
parameters; \citealt{gustafson2015bayesian}), that point estimates
will be unbiased and the 90\% credible intervals will cover the true
value of the parameters with a probability of 0.90.

For all settings, our simulation experiment shows that point estimates
of the time and location of introduction are apparently unbiased (Figs
S8\textendash S9 in Appendix S3) and the 90\% credible intervals cover
the true location and time with a probability of $\geq0.86$. Appendix
S3 contains details of our simulation experiment, full results, and
computer code needed to reproduce our experiment.

\subsection{Forecast Comparison}

Accurate forecasts during the early stages of disease outbreak or
invasion of a novel species may be useful to inform management recommendations
and surveillance sampling. In addition to estimating the location
and date of introduction, an auxiliary benefit of Bayesian implementations
of dynamic models is that probabilistic predictions and forecasts
can be obtained as a derived quantity (\citealt{Hobbs2015,hefley2017mechanism}).
To demonstrate the accuracy of forecasts from our model, we used the
samples collected from 2012 that were reserved for this single purpose
(see WNS data description for more detail). To facilitate comparison
of our model with other non-Bayesian approaches, we calculated the
zero-one score using the posterior mean of the probability of infection
(i.e., $\text{E}(p_{i}|\mathbf{y})$) and forecasted that an individual
sample would be WNS ``positive'' if $\text{E}(p_{i}|\mathbf{y})\geq0.5$
and ``negative'' if $\text{E}(p_{i}|\mathbf{y})<0.5$ (\citealt{gneiting2007strictly,gneiting2011making}).
Although any local and proper scoring rule could be used, we chose
the zero-one score because it can be used to calculate the misclassification
rate, which is easy to interpret and communicate to practitioners.
The misclassification rate is calculated using the 2012 data by summing
the total number of samples incorrectly classified and dividing by
the total number of samples.

As a standard of comparison, we followed \citet{hefley2017mechanism}
and compared the forecasting ability of our model to phenomenological
regression-type models developed for classification of binary events
(e.g., positive/negative) from the statistics and machine learning
literature (\citealt{james2013introduction}). Specifically, we compared
forecasts from our model to three approaches, which include: (1) a
generalized linear model (GLM) that included deciduous forest cover
and karst terrain at the sample location and bat species as predictor
variables; (2) a generalized additive model (GAM) that included a
linear effect of the same predictor variables as the GLM, but that
accounted for spatial autocorrelation by using two-dimensional splines
on a sphere (\citealt{wood2017generalized}); and (3) the same GAM
specification as in 2, but with an additional smooth effect of time
to account for increasing prevalence and temporal autocorrelation.
Details associated with the exact model specifications and computations
are reported in Appendix S2.

For each approach, we used the samples collected from 2012 and calculated
the zero-one score using the maximum (or restricted maximum) likelihood
estimate of the predicted probability of infection ($\hat{p}_{i}$)
and forecasted that an individual sample would be WNS ``positive''
if $\hat{p}_{i}\geq0.5$ and ``negative'' if $\hat{p}_{i}<0.5$.
To compare forecasts with our model, we report the misclassification
rate for each approach. Although comparing forecasts among different
models is useful for quantifying forecast accuracy, this comparison
is does not explicitly evaluate the reliability of the inferred time
and location of introduction.

\section{Results}

The year with the highest probability of introduction was 2002 (1992\textendash 2005;
90\% credible interval), which is 4 years before the first photographic
evidence documented the presence of $\textit{P. destructans}$ in
the United States (Fig. 4). When compared to the location of first
detection based on photographic evidence from 2006, the posterior
distribution shows an area covering $123,798$ km$^{2}$ that has
an equal or higher probability of containing the point where $\textit{P. destructans}$
was introduced (Fig. 4). This $123,798$ km$^{2}$ area represents
a 74\% credible region and intersects the location of first detection,
but indicates that location of introduction that are equally or more
probable could be up to 510 km east of the first detection. For comparison,
the 90\% posterior credible region covers an area of $182,493$ km$^{2}$.
This $182,493$ km$^{2}$ regions represents 2.3\% of the area covered
by the 90\% prior credible region, which demonstrates a large reduction
in uncertainty due to the WNS surveillance data.
\begin{figure}[H]
\begin{centering}
\includegraphics[scale=0.65]{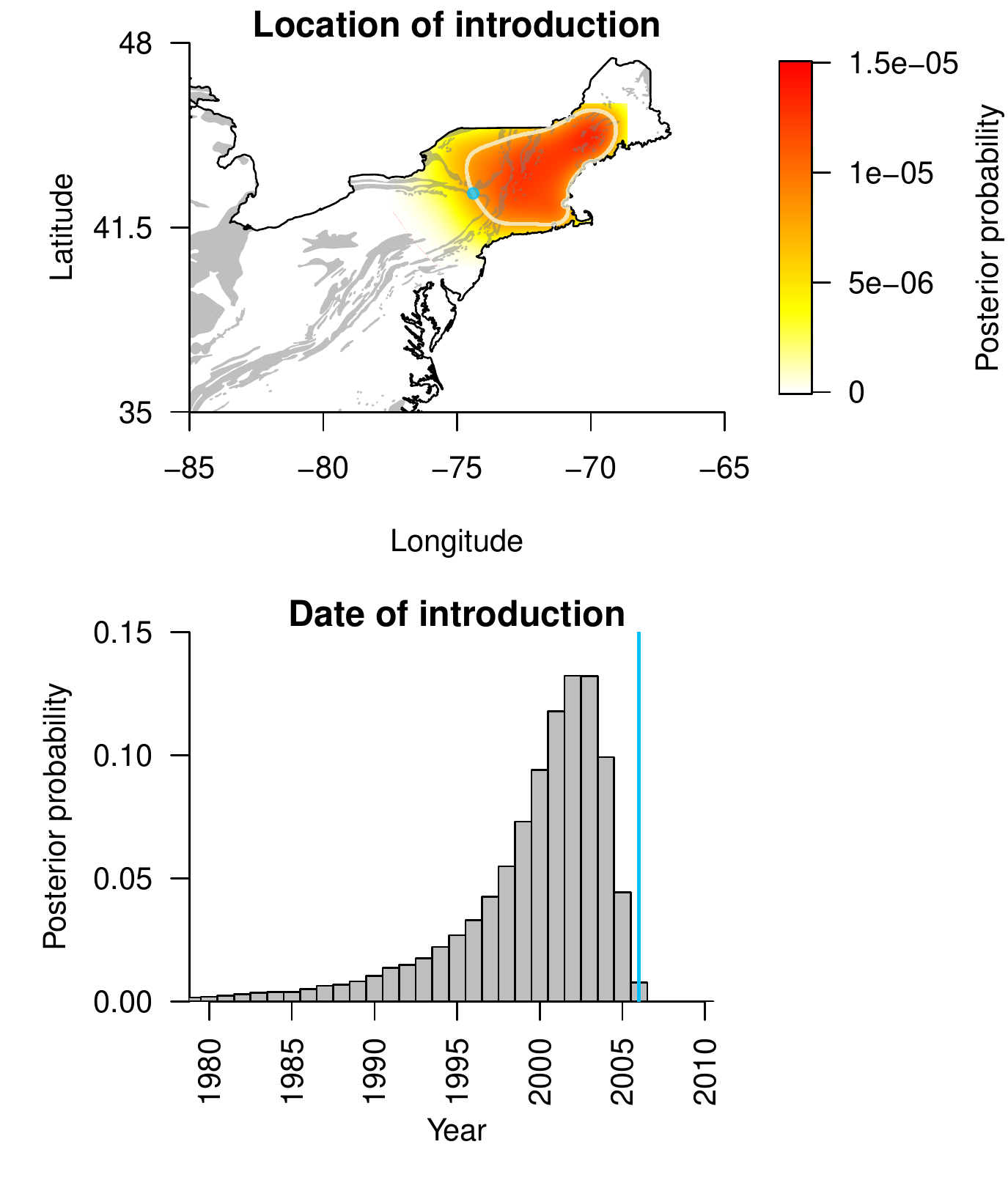}
\par\end{centering}
\noindent \textbf{Fig. 4.} Posterior distributions of the location
and date of introduction obtained from fitting our model to the data
shown in Fig. 1. The top panel shows the joint posterior distribution
of the location of introduction with red indicating a higher probability
than yellow. The beige polygon outlines an area that is equally or
more likely to contain the point of introduction when compared to
the location where white-nose syndrome was first detected in 2006
(blue dot). The gray areas are karst formations with caves, which
is likely winter bat habitat. The bottom panel shows the marginal
posterior distribution of the date of introduction. The blue vertical
line represents the earliest known photographic evidence of white-nose
syndrome on bats in the United States. Years with low probability
are not shown (i.e., before 1980).
\end{figure}

In addition to estimating the location and date of introduction, our
approach simultaneously estimates species-specific susceptibility
(Fig. 5), diffusion and growth rates of the pathogen (Fig. 6), and
can be used to forecast the spread of $\textit{P. destructans}$ (Fig.
7 and animation in Appendix S4). As it relates to species-specific
susceptibility, our results show that the northern long-eared bat
was the most susceptible to $\textit{P. destructans}$ whereas the
big brown bat was the least susceptible (Fig. 5). The diffusion and
growth rates are positive in many regions and show spatial variability
that is attributable to deciduous forest cover and karst terrain that
contains caves (cf., Fig. 3 and Fig. 6, Fig. S3). As a result of positive
growth and diffusion rates, forecasts show that $\textit{P. destructans}$
spread rapidly across the contiguous United States (Fig. 7), causing
an overall increase in the probability of infection for all four species
of bats over time (see animations in Appendix S4).

Our forecast comparison revealed a misclassification rate of 26\%
for our hierarchical Bayesian model. For comparisons, the three alternative
approaches we used resulted in misclassification rate of 43\% for
method one (GLM), 28\% for method 2 (GAM that included karst, forest
and species as predictors and that accounted for spatial autocorrelation)
and 30\% for method three (same as method 2, but accounted for temporal
dynamics and autocorrelation). A visual comparison of the forecasts
for all methods is given in Appendix S2 (Fig. S5).
\begin{figure}[H]
\begin{centering}
\includegraphics[scale=0.55]{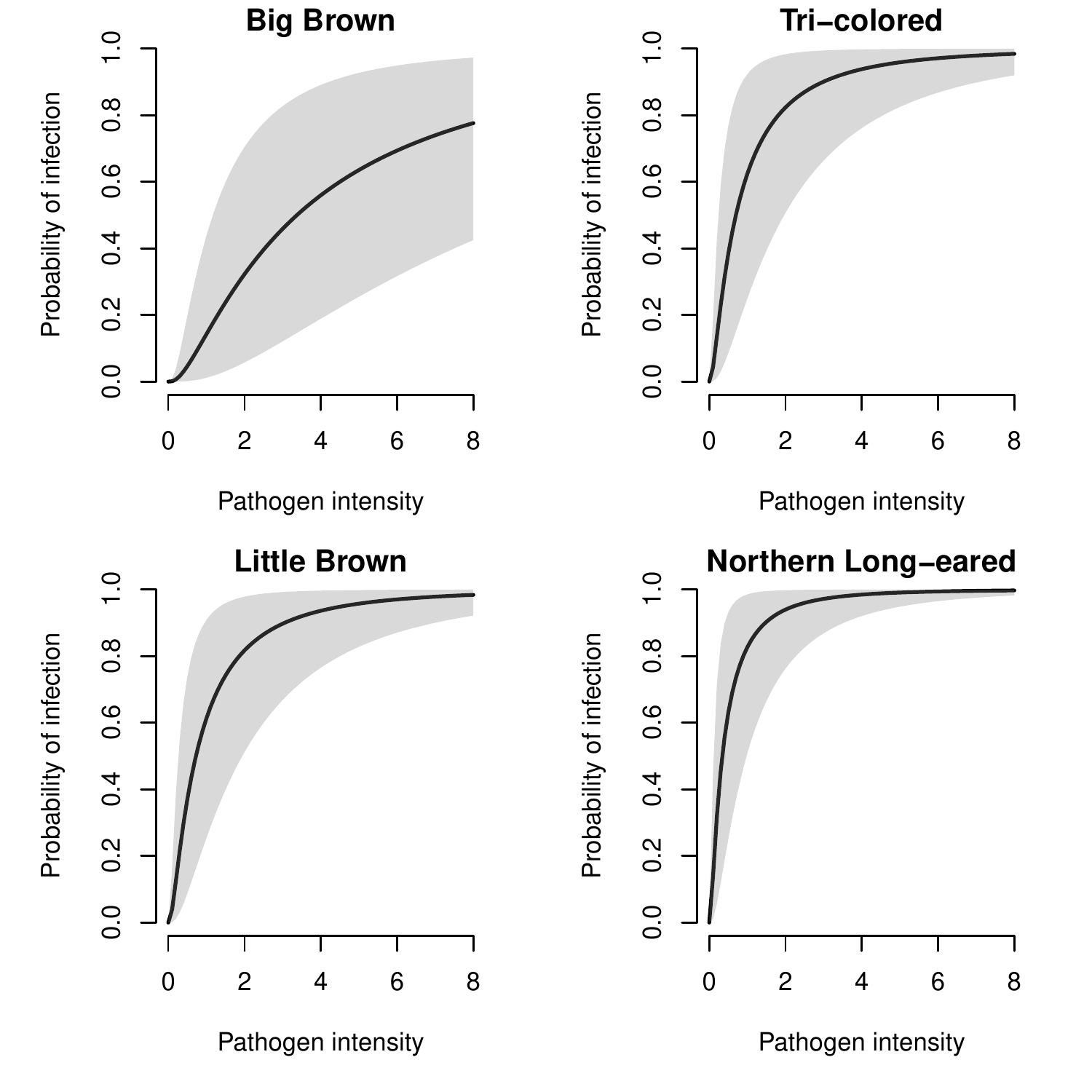}
\par\end{centering}
\noindent \textbf{Fig. 5.} The estimated species-specific susceptibility
depends on the intensity of $\textit{P. destructans}$ ($u(\mathbf{s},t)$).
Shown is the expected value of the posterior distribution (black lines)
and 95\% credible intervals (gray shading).
\end{figure}
\begin{figure}[H]
\begin{centering}
\includegraphics[scale=0.9]{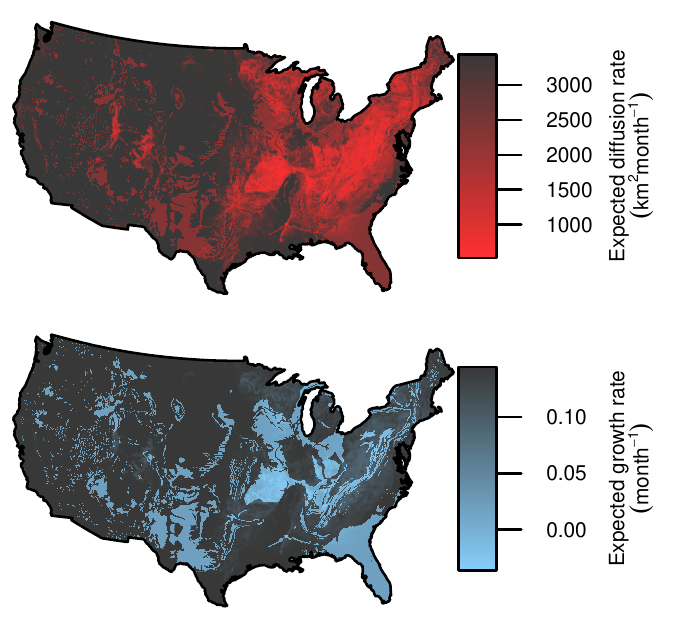}
\par\end{centering}
\noindent \textbf{Fig. 6.} The estimated diffusion rate (top panel)
determines how fast the pathogen $\textit{P. destructans}$ will spread,
whereas the growth rate (bottom panel) determines how quickly the
intensity of $\textit{P. destructans}$ will increase. Deciduous forest
and karst terrain (Fig. 3) influence the posterior distributions of
the diffusion and growth rate leading to the spatial variability appearing
in the maps.
\end{figure}
\begin{figure}[H]
\begin{centering}
\includegraphics[scale=0.7]{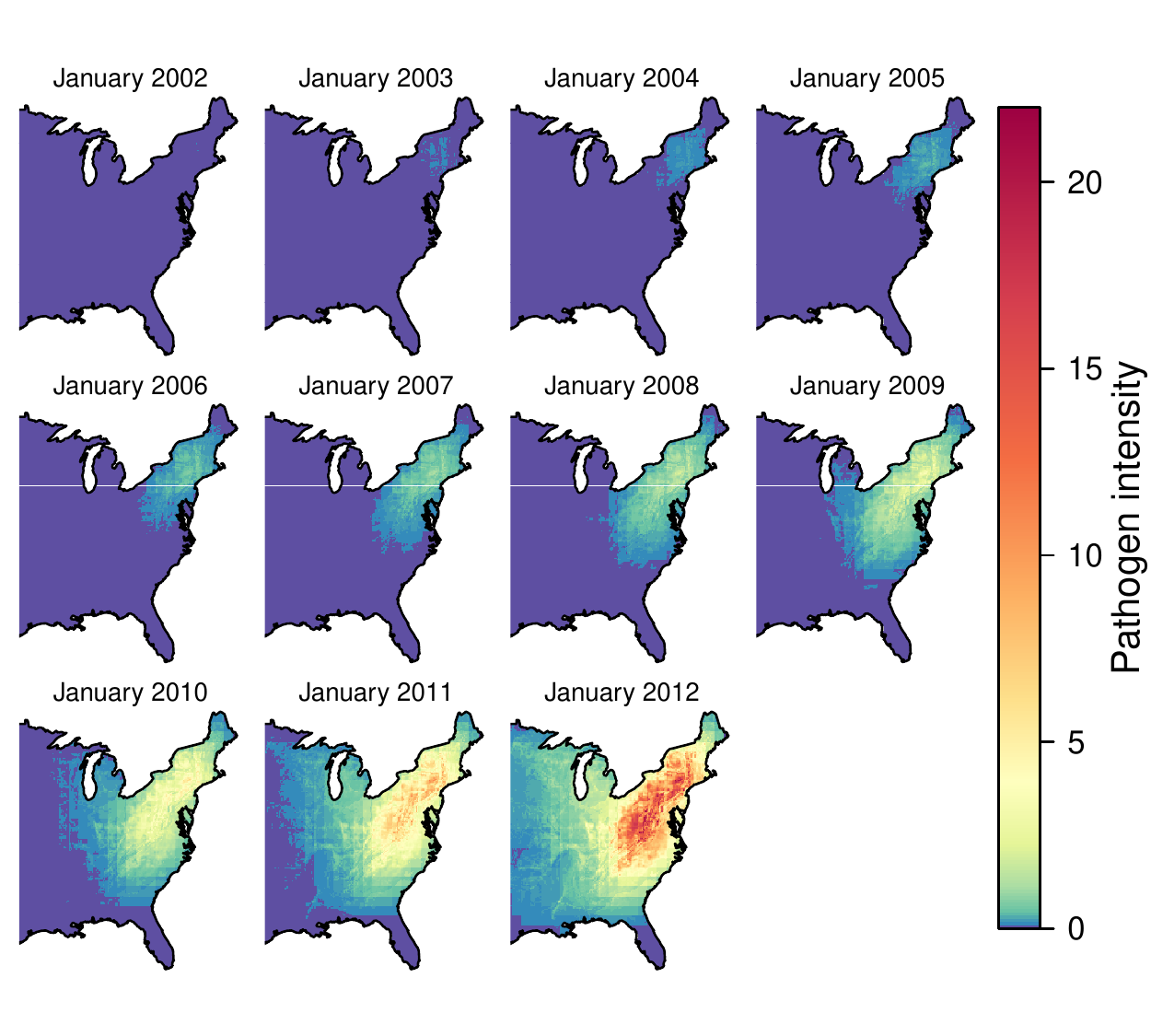}
\par\end{centering}
\noindent \textbf{Fig. 7.} Estimated intensity of $\textit{P. destructans}$
($u(\mathbf{s},t)$) for January 2002\textendash 2012. See Appendix
S4 for animations showing the contiguous United States for all months
over the period 2000\textendash 2012 as well as forecasts of the probability
of infection for all four species of bats. Note that the pathogen
intensity, $u(\mathbf{s},t)$, can be linked to the expected pathogen
abundance by integrating $u(\mathbf{s},t)$ over a spatial domain
of interest (i.e., $\int_{\mathcal{A}}u(\mathbf{s},t)d\mathbf{s}$).
\end{figure}

\section{Discussion}

Our results reveal that it is possible to estimate the date and location
of introduction shortly after the release of a pathogen using indirect
data from 397 samples collected over 4 years. Our results also demonstrate
the need for statistical thinking because the estimated date shows
that the introduction likely occurred several years before the first
detected case. Conversely our results show that the location of first
detection is a plausible location of introduction, but there is a
moderate level of uncertainty surrounding the exact location.

For WNS, the estimated location and date of introduction, along with
the associated uncertainty, constitutes knowledge generated by harnessing
the information within disease surveillance data using statistically
valid techniques. Similar surveillance data is routinely collected
for diseases of wildlife, plants, livestock, and humans, but filling
knowledge gaps using these data requires statistical methods that
capture dynamic processes and quantify uncertainty.

The introduction of $\textit{P. destructans}$ is suspected to have
originated from a single point source due to the clonal nature of
the fungal pathogen in North America (\citealt{drees2017phylogenetics}).
For other pathogens or pests, introductions may occur at multiple
locations. When multiple introductions occur, we explain in Appendix
S1 how our approach can be modified accordingly and potential issues
that may arise when fitting the model to data. When multiple introduction
occur, our approach could identify areas and environmental conditions
where new pathogens or pest introductions are likely to occur\textendash effectively
providing a map that is similar to Fig. 4, but showing areas at high
risk for new introductions rather than a single past introduction.
With this information, new introductions could be anticipated and
perhaps avoided or optimal monitoring and control programs could be
established (\citealt{williams2017monitoring}; \citealt{laber2018optimal}).

Although our study system was epidemiological, dynamic statistical
models have been used to explain the outbreak of pests (\citealt{powell2014phenology,hooten2013computationally,powell2017differential}),
the invasions of a species (\citealt{neupane2015invasion}), the reintroduction
of an extirpated species (\citealt{williams2017integrated}), and
to determine optimal monitoring of ecological processes (\citealt{williams2017monitoring}).
Such a range of applications indicate that dynamic statistical models
can be used for a variety of situations including to determine the
location of remnant populations that enable outbreaks of pests, estimate
the date, location, and propagule size where invasive species were
introduced, or determine the habitat characteristics where an extirpated
species can re-establish.

\section*{\vspace{-0.5cm}
Acknowledgements}

We thank all state, federal and other partners for submitting samples
and the U.S. Geological Survey National Wildlife Health Center for
processing the samples. At the time of publication data were not available
for release because of the sensitivity related to the locations of
the bat species sampled. Data can be requested from the following
agencies that submitted samples. Agencies that submitted samples include
Acadia National Park, Aerospace Testing Alliance Conservation, Alabama
Department of Conservation and Natural Resources, Alabama Division
of Wildlife and Freshwater Fisheries, Alaska Department of Fish and
Game, U.S. Fish and Wildlife Service (USFWS) Alligator River National
Wildlife Refuge, Arizona Game and Fish Department, Arkansas Department
of Health, Arkansas Game and Fish Commission, U.S. Air Force, Avon
Park Air Force Range, Bat Conservation International, Bat World Sanctuary,
U.S. Department of Agriculture (USDA) Forest Service, Black Hills
National Forest, Bureau of Land Management, Boston University, Buffalo
National River (National Park Service), California Department of Fish
and Game, Canadian Cooperative Wildlife Health Centre, Channel Islands
National Park, Clinton Wildlife Management Area, Colorado Department
of Natural Resources, USDA Forest Service Colville National Forest,
Connecticut Department of Environmental Protection, Delaware Department
of Natural Resources and Environmental Control, Department of the
Army, Mississippi Department of Wildlife, Fisheries and Parks, Devils
Tower National Monument, East Stroudsburg University, Ecological Sciences,
Inc., Florida Fish and Wildlife Conservation Commission, Frenso Chaffee
Zoo, Furman University, Georgia Department of Natural Resources, USDA
Forest Service Gifford Pinchot National Forest, Great Smoky Mountains
National Park, USDA Forest Service Hoosier National Forest, Humboldt
State University, Idaho Department of Fish and Game, Illinois Department
of Natural Resources, Illinois Natural History Survey, Indiana State
University, Iowa Department of Natural Resources, U.S. Air Force,
Langley, Kansas Department of Parks, Wildlife, and Tourism, Lake of
the Ozarks State Park, Lake Vermilion-Soudan Underground Mine State
Park, Lava Beds National Monument, USDA Forest Service, Lincoln National
Forest, Louisiana Department Baton Rouge, Maine Department of Inland
Fisheries and Wildlife, Maryland Department of Agriculture, Maryland
Department of Natural Resources, Massachusetts Division of Fish and
Wildlife, Michigan Department of Natural Resources, Ministere des
Ressources naturelles et de la Faune, Minnesota Department of Natural
Resources, Missouri Department of Conservation, Missouri State University,
Montana Department of Fish, Wildlife and Parks, Nature Conservancy,
Nebraska Game and Parks Commission, Nevada Department of Wildlife,
New Hampshire Fish and Game Commission, New York Department of Environmental
Conservation, North Carolina Wildlife Resources Commission, North
Dakota Game and Fish, North Mississippi Refuges Complex, USFWS Noxubee
National Wildlife Refuge, EL Malpais National Monument, Ohio Department
of Natural Resources, Ohio Wildlife Center, Oklahoma State University,
Oklahoma Department of Wildlife Conservation, Olentangy Wildlife Research
Station, Oregon Caves National Monument, Oregon Department of Fish
and Wildlife, Oregon Parks and Recreation Department, Ozark National
Scenic Riverway, Ozark Plateau National Wildlife Refuge, Pennsylvania
Game Commission, Pennsylvania State Animal Diagnostics Laboratory,
Pima County Natural Resources, Parks, and Recreation, Rhode Island
Department of Environmental Management, Rhode Island Division of Fish
and Wildlife, USFWS San Bernardino/Leslie Canyon National Wildlife
Refuge, South Dakota Department of Game, Fish and Parks, Session Woods
Wildlife Management Area, USDA Forest Service, Shawnee National Forest,
Sleeping Bear Dunes National Lakeshore, South Carolina Department
of Natural Resources, South Mountain Wildlife Rehabilitation Center,
Tennessee Wildlife Resources Agency, Texas Parakeet Raisers, Texas
Parks and Wildlife, University of Central Oklahoma, University of
Montreal, University of Nebraska, USFWS Upper Mississippi River National
Wildlife and Fish Refuge, USDA Forest Service National Forests and
Grasslands in Texas, USDA Forest Service Hot Springs Arkansas, USDA
Animal and Plant Health Inspection Service Wildlife Services, USDA
Forest Service North Central Research Station, USDA Forest Service
Pacific Southwest Research Station, USFWS Arkansas Field Office, USFWS
Ecological Services, USFWS Environmental Contaminants, Utah Division
of Wildlife Resources, Vermont Fish and Wildlife, Virginia Department
of Conservation and Recreation, Virginia Division of Natural Heritage,
Virginia Game and Inland Fisheries, Washington Department of Fish
and Wildlife, West Virginia Department of Natural Resources, USFWS
Wheeler National Wildlife Refuge, USDA Forest Service White River
National Forest, Wisconsin Dept of Natural Resources, Wisconsin State
Lab of Hygiene, Wyoming Fish and Game. We thank Cydney Alexis, Nora
Bello, Hannah Birg$\acute{\text{e}}$, Jeffrey Lorch, Bob O'Hara,
Daniel Walsh, and Perry Williams for insightful discussions and comments
that improved this manuscript. The authors acknowledge support for
this research from USGS G16AC00413. Any use of trade, firm, or product
names is for descriptive purposes only and does not imply endorsement
by the U.S. Government.\vspace{-0.5cm}

\section*{Supplementary Material}

\noindent \textbf{Appendix S1}

\noindent Tutorial with R code to simulated data and fit the Bayesian
ecological diffusion model.

\noindent \textbf{Appendix S2}

\noindent Tutorial and R code to reproduce the white-nose syndrome
data analysis.

\noindent \textbf{Appendix S3}

\noindent Simulation experiment description, results, and R code.

\noindent \textbf{Appendix S4}

\noindent Supplementary animations of Fig. 7.

\vspace{-0.5cm}

\begin{singlespace}
\noindent \bibliographystyle{apa}
\bibliography{references}

\end{singlespace}

\end{spacing}
\end{flushleft}
\end{document}